\documentclass[twocolumn]{aastex631}

\shorttitle{Kepler-139f: discovery and dynamics}
\shortauthors{Lammers \& Winn}

\graphicspath{{./}{}}
\usepackage{amsmath}
\usepackage{comment}

\begin{document}

\newcommand{\CL}[1]{\textcolor{purple}{CL: #1}}

\title{Discovery and Dynamics of the Nontransiting Planet Kepler-139f}


\author[0000-0001-9985-0643]{Caleb Lammers}
\affiliation{Department of Astrophysical Sciences, Princeton University, 4 Ivy Lane, Princeton, NJ 08544, USA}

\author[0000-0002-4265-047X]{Joshua N.\ Winn}
\affiliation{Department of Astrophysical Sciences, Princeton University, 4 Ivy Lane, Princeton, NJ 08544, USA}

\begin{abstract}

Among the ways that an outer giant planet can alter the architecture of an inner planetary system is by tilting the orbits of the inner planets and reducing their mutual transit probabilities. Here, we report on an example of this phenomenon: we show that the Kepler-139 system contains a nontransiting planet just exterior to three transiting planets, and interior to a giant planet. This newly discovered planet, Kepler-139f, has an orbital period of $355\,{\pm}\,2$\,days and a mass of $36\,{\pm}\,10\,M_\oplus$ based on transit-timing and radial-velocity data. Through dynamical simulations, we show that gravitational perturbations on planet f's orbit from the outer giant planet reduce the probability for a randomly located observer to see transits of all four inner planets. Thus, Kepler-139 illustrates the role that outer giant planets can play in the apparent truncation of compact systems of multiple transiting planets.
\end{abstract}

\section{Introduction}
\label{sec:intro}

Thanks to NASA's {\it Kepler} survey for transiting planets, we know that Sun-like stars often host systems of multiple sub-Neptune-sized planets with orbital periods shorter than a few hundred days \citep{Borucki2011, Fressin2013, Zhu2018}. These ``compact multiplanet systems'' typically feature planets with similar sizes and masses \citep{Lissauer2011a, Weiss2018,Millholland2017, Otegi2022}. Their orbits have low mutual inclinations (${\sim}\,1^\circ$; \citealt{Lissauer2011a, Fang&Margot2012, Fabrycky2014}) and low eccentricities (${\lesssim}\,0.05$; \citealt{VanEylen&Albrecht2015, Hadden&Lithwick2017}), with spacings that form roughly geometric progressions \citep{Weiss2018, Gilbert&Fabrycky2020}. Despite these advances in our knowledge of the architectures of compact multiplanet systems, many unanswered questions remain. The question most relevant to this paper is: how many planets go unseen because they do not transit?

Transit surveys are subject to strong observational biases \citep[see, e.g.,][]{Pepper+2003, WinnPetigura2024}. Planets inclined by more than a few degrees with respect to our line of sight are often missed, making {\it Kepler}'s census of planetary systems incomplete. Radial velocity (RV) observations have helped to complete the picture, especially by providing sensitivity to massive, far-out planets. RV monitoring of {\it Kepler} systems has been undertaken by several groups to seek any associations between the existence of compact multiplanet systems and the presence of outer giant planets \citep{Zhu&Wu2018, Bryan2019, Rosenthal2022, Bonomo2023, Bryan&Lee2024}.

Outer giant planets can exert profound dynamical influences on the architectures of inner planetary systems. The ability of outer giants to dynamically excite their small inner companions has been invoked to explain at least three observed trends in the demographics of transiting planets. First, dynamical disruption due to outer giants has been proposed as an explanation for the relatively large fraction of {\it Kepler} systems that have only one known transiting planet \citep{Lai&Pu2017, Huang2017, Read2017}. Second, outer giants might be responsible for truncating compact multiplanet systems to have maximum orbital periods of several hundred days \citep{Millholland2022, Sobski&Millholland2023}. Third, dynamical excitation has been invoked to explain unusually large gaps between orbits of neighboring planets in systems that host outer giants (\citealt{He&Weiss2023, Livesey&Becker2024}).

One way to test these hypotheses is to search for nontransiting planets in systems that host transiting planets. Unfortunately, finding nontransiting planets of the expected size -- smaller than Neptune -- has proven challenging. For a handful of systems, it has been possible to use transit timing variations (TTVs) to infer the existence of a nontransiting planet \citep[e.g.,][]{Ballard2011, Nesvorny2012}. The NASA Exoplanet Archive\footnote{\url{https://exoplanetarchive.ipac.caltech.edu/} (accessed on March 10th, 2025)} reports only four cases in which TTVs have been used to discover a nontransiting planet with a mass below $100\,M_\oplus$ and a well-determined orbital period. Those four planets are Kepler-19c \citep{Malavolta2017}, Kepler-82f \citep{Freudenthal2019}, Kepler-411e \citep{Sun2019}, and Kepler-138e \citep{Piaulet2023}. None of these systems are known to host an outer giant planet.

Motivated by the theoretical expectation that outer giants will at least occasionally disrupt their inner systems, we have been searching for nontransiting planets in the subset of {\it Kepler} systems known to harbor wide-orbiting giant planets. Here, we report the discovery of a ${\sim}\,35$-$M_\oplus$ nontransiting planet around Kepler-139, a G-type star
with a $V$-band magnitude of $12.7$ that hosts three transiting planets and an RV-detected outer giant. The transiting plants have orbital periods $P_d\,{=}\,7.31$\,days, $P_b\,{=}\,15.8$\,days, and $P_c\,{=}\,157$\,days and radii $R_d\,{=}\,1.7\,R_\oplus$, $R_b\,{=}\,2.4\,R_\oplus$, and $R_c\,{=}\,2.5\,R_\oplus$ (\citealt{Fulton&Petigura2018}; note that the innermost planet is named ``d'' not ``b''). The outer giant has an orbital period of $P_e\,{\approx}\,2{,}000$\,days and a mass of $m_e\,{\approx}\,400\,M_\oplus$. Below, we present the TTV and RV evidence for the new planet and discuss the relevance of the outer giant to the nontransiting orbital orientation of Kepler-139f.

\section{Measuring {\it Kepler} transit times}
\label{sec:transit_fitting}

Transit times for Kepler-139d, b, and c were reported by \citet{Holczer2016} as part of their large catalog of TTVs for {\it Kepler} objects of interest. To improve the uncertainty estimates and remove timing outliers, we have re-analyzed Kepler-139's {\it Kepler} light curves. With the help of the \texttt{lightkurve} package \citep{Lightkurve2018}, we downloaded all available {\it Kepler} photometry from the Mikulski Archive for Space Telescopes (MAST). We used the short-cadence data (1-minute averaging) whenever it was available; otherwise, we used the long-cadence data (30-minute averaging). The resulting dataset contains $182$, $87$, and $9$ transits of planets d, b, and c, respectively.

\begin{figure}
\centering
\includegraphics[width=0.475\textwidth]{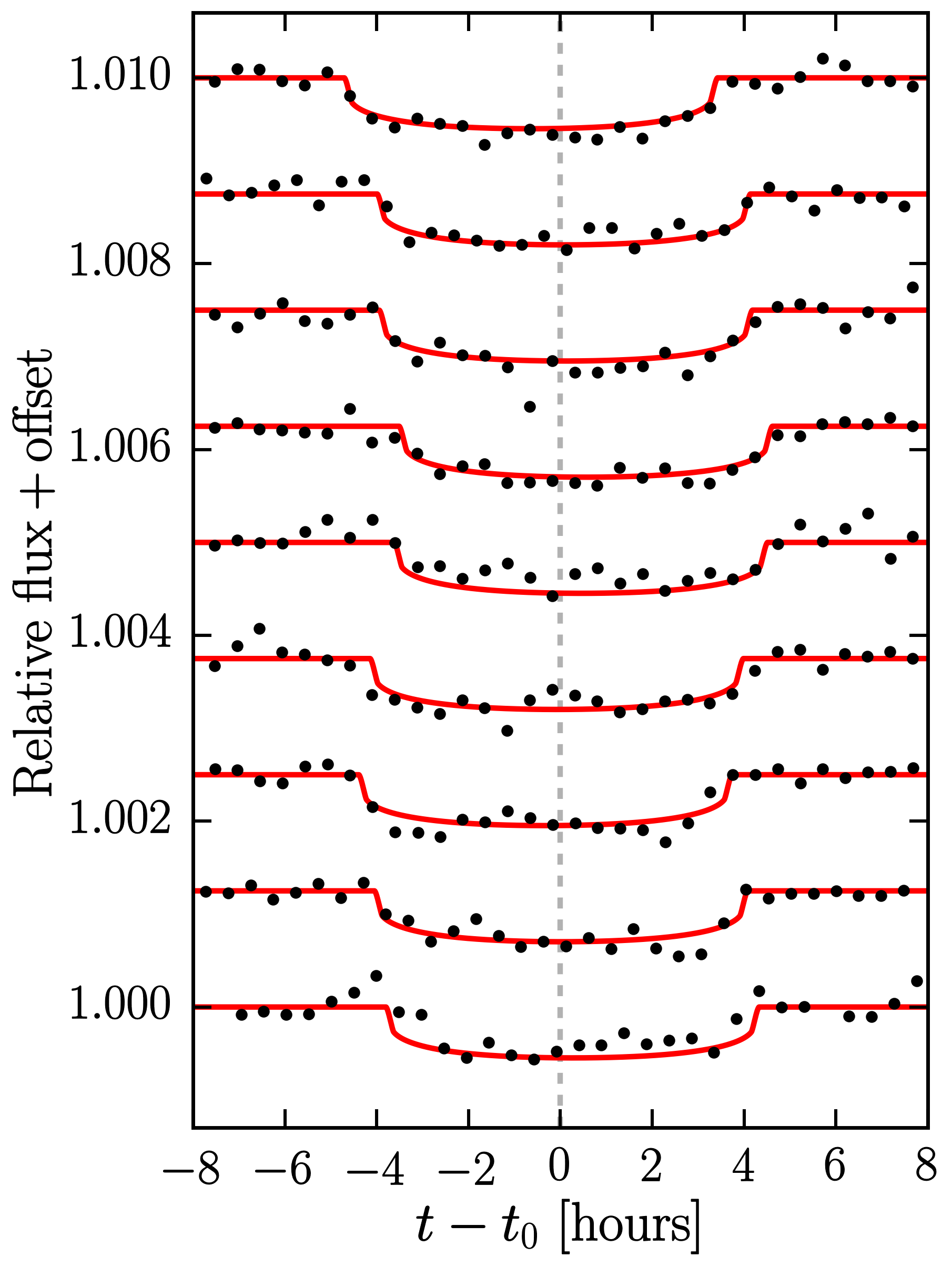}
\caption{Transit light curves of Kepler-139c from {\it Kepler} (black) along with the best-fit models (red). Fits were performed using the short cadence (one-minute) light curves when available, but the data are shown here in 30-minute bins for ease of visualization. The gray dashed line marks the expected transit midpoint if the planet's period were exactly constant. The data reveal transit-timing variations of up to ${\sim}\,0.5$\,hour.}
\label{fig:tlc_fits}
\end{figure}

To measure the transit times, we fitted the data spanning each observed transit with the standard \citet{Mandel&Agol2002} model. Before doing so, we removed photometric trends in the data by fitting a polynomial function of time to the out-of-transit data over a time interval centered on the predicted transit midpoint and lasting three times the predicted transit duration (with predictions based on the tables of \citealt{Thompson2018}). The degree of the polynomial was 1, 2, or 3, chosen by minimizing the Bayesian Information Criterion (BIC), $\mathrm{BIC}\,{=}\,\chi^2\,{+}\,k\ln(n)$, where $k$ is the number of free parameters in the model and $n$ is the number of data points \citep{Schwarz1978}. A normalized light curve was produced by dividing the flux by the lowest-BIC polynomial. Because individual transits were detected with a relatively low signal-to-noise ratio, we chose to fix the transit parameters for each planet ($R_p/R_\ast$, $a/R_\ast$, and $b$) to the values tabulated by \cite{Thompson2018}. We also assumed a quadratic limb-darkening law with coefficients $u_1\,{=}\,0.44$ and $u_2\,{=}\,0.24$, based on the tables of \citet{Claret&Bloemen} for a star with $T_\mathrm{eff}\,{=}\,5680$\,K, $\mathrm{[Fe/H]}\,{=}\,0.28$, and $\log(g)\,{=}\,4.35$ \citep{Morton2016}.\footnote{We interpolated the \citet{Claret&Bloemen} table using the online tool of \citet{Eastman2013}, which is available at \url{https://astroutils.astronomy.osu.edu/exofast/limbdark.shtml}.} We determined the best-fit mid-transit time by minimizing $\chi^2$ with the Nelder-Mead optimizer from \texttt{SciPy}'s \texttt{optimize.minimize} class \citep{Virtanen2020}. To determine the uncertainties in the measured transit times, we performed an affine-invariant Markov Chain Monte Carlo (MCMC) analysis \citep{Goodman&Weare2010} with the help of the \texttt{emcee} code \citep{Foreman-Mackey2013}. We used $100$ independent walkers which each took $10{,}000$ steps, the first $5{,}000$ of which we discarded as burn-in. The resulting Markov chains were ${>}\,50$ times longer than the autocorrelation length for all but a few cases. The typical timing uncertainty was ${\sim}\,5$\,min. We discarded transit times with anomalously low or high formal uncertainties (${<}\,1$\,min or ${>}\,20$\,min), which corresponded to partial transits or unusually noisy light curves. This left us with $142$ transit times for planet d, $85$ for planet b, and $9$ for planet c. Figure~\ref{fig:tlc_fits} shows the results for Kepler-139c, for which the TTVs are large enough to be seen by eye.

\begin{figure*}
\centering
\includegraphics[width=0.975\textwidth]{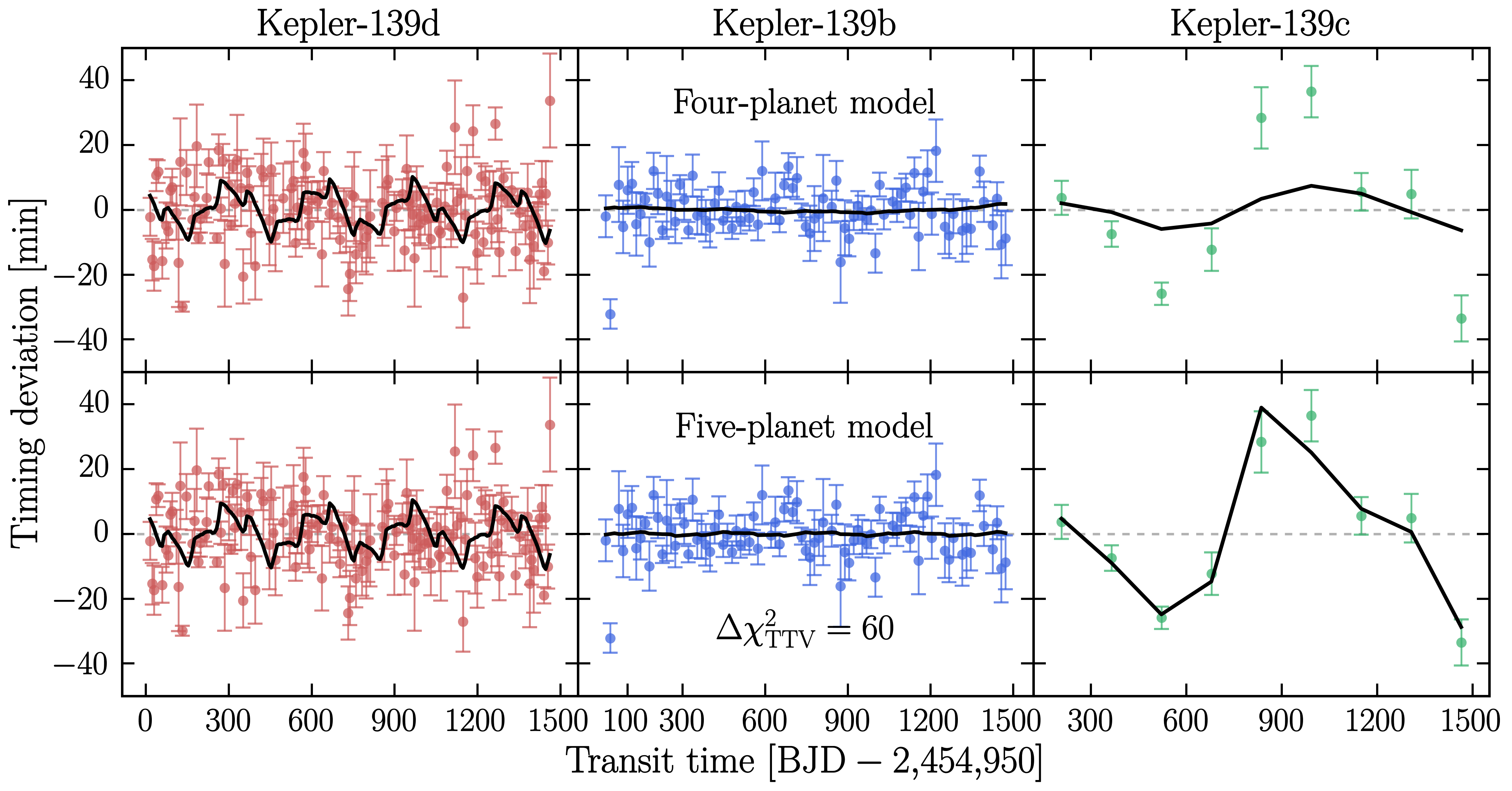}
\caption{TTVs measured for Kepler-139d (red), Kepler-139b (blue), and Kepler-139c (green). In the top row, the black curves are based on a model fitted to the TTVs and RVs that includes only the four previously known planets. The four-planet model fails to reproduce the $30$-minute TTVs of planet c. In the bottom row, the black curves are based on a five-planet model, including a $45$-$M_\oplus$ nontransiting planet with a period of $354$ days, placing it wide of the 2:1 MMR with planet c. The five-planet model provides a far superior fit ($\Delta\chi_\mathrm{TTV}^2\,{=}\,60$).}
\label{fig:TTVs}
\end{figure*}

\section{TTV and RV joint modeling}
\label{sec:TTV_model}

We modeled the TTVs using $N$-body simulations with the help of the \texttt{TTVFast} code \citep{Deck2014}.\footnote{Specifically, we used the Python wrapper of \texttt{TTVFast} available at \url{https://github.com/simonrw/ttvfast-python}.} \texttt{TTVFast} uses a \citet{Wisdom&Holman1991} integrator to advance the positions of the planets and repeatedly checks for transits by tracking the planets' sky-projected star-planet distances. When a transit is detected, the code refines the transit time calculation by assuming the motion to be Keplerian and employing Newton's method to achieve an accuracy better than 10~sec.

We parameterized the Kepler-139 system using five parameters for each planet: the mass $m$, the orbital period $P$, the two eccentricity vector components $k\,{=}\,e\cos(\varpi)$ and $h\,{=}\,e\sin(\varpi)$, and the initial mean longitude $\lambda$. The initial condition was specified at an arbitrarily chosen reference time of $T_\mathrm{epoch}\,{=}\,2{,}454{,}950$\,BJD. We assumed all the planets have coplanar orbits ($i\,{=}\,90^\circ$), after confirming through numerical experiments that mutual inclinations of a few degrees do not affect the calculated TTVs to within the observational uncertainties \citep[see also][]{Hadden&Lithwick2016}. We chose a Wisdom-Holman time step of $P_1/20\,{\approx}\,0.365$\,days and set the stellar mass to be $M_\ast\,{=}\,1.078\,M_\odot$ (from \citealt{Fulton&Petigura2018}).

Kepler-139 has undergone RV monitoring between $2010$ and $2022$ as a part of the Kepler Giant Planet Survey (KGPS; \citealt{Weiss2024}). Over the past 12 years, $38$ RVs were collected using the W.\ M.\ Keck Observatory High Resolution Echelle Spectrometer, with a formal uncertainty of about $2$\,m\,s$^{-1}$ \citep{Weiss2024}. We fitted the RVs and TTVs jointly with our \texttt{TTVFast}-based model, after adding another free parameter to represent the arbitrary RV offset.

\begin{figure*}
\centering
\includegraphics[width=0.95\textwidth]{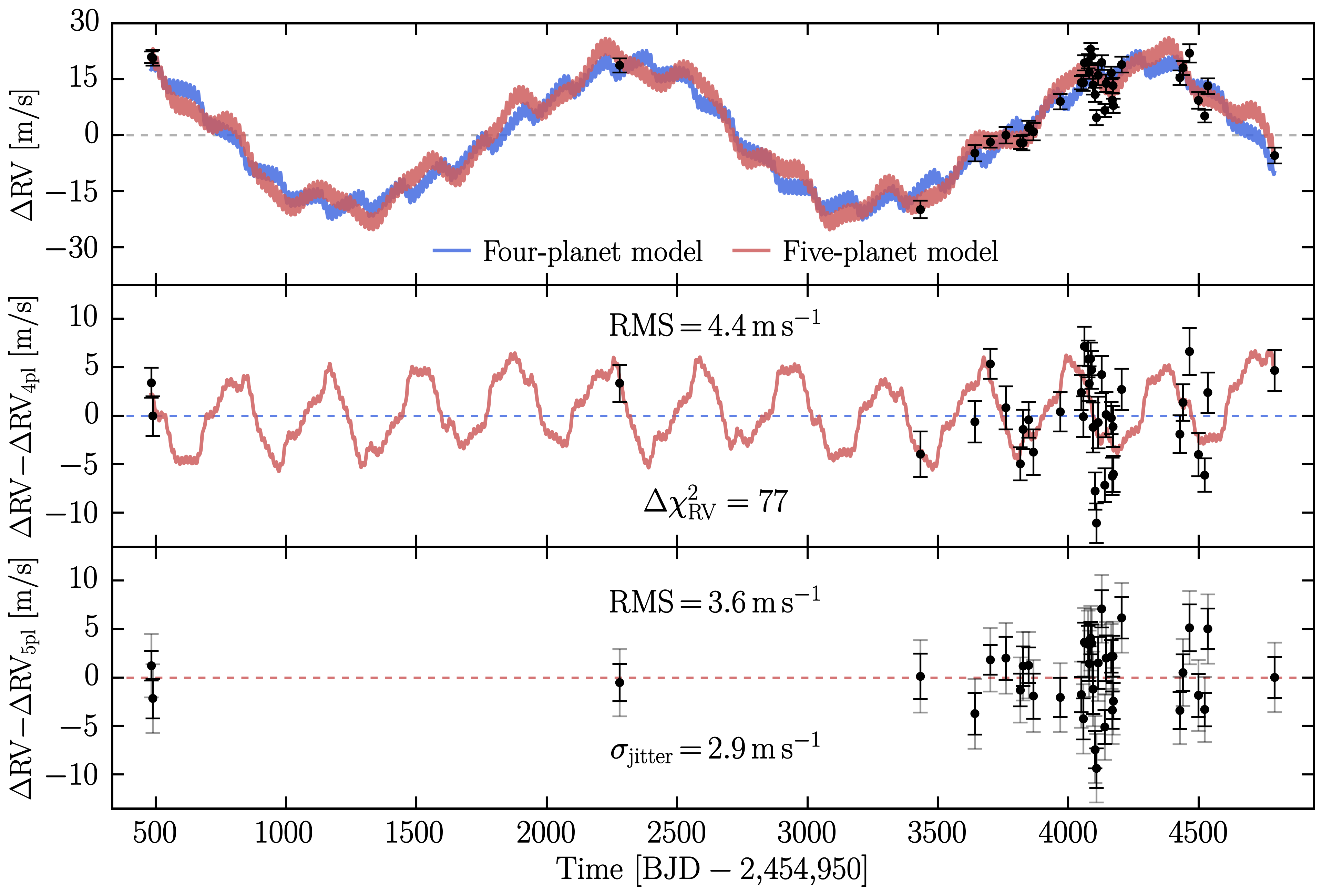}
\caption{RV data for Kepler-139 (black) from \citet{Weiss2024}. The blue curve is the four-planet model that best fits the RV and TTV data (residual root-mean-square of $4.4$\,m\,s$^{-1}$). The red curve is the best-fit five-planet model ($3.6$\,m\,s$^{-1}$ residual root-mean-square). The middle panel shows the deviations between the data and the best-fit four-planet model, and the bottom panel shows the same for the five-planet model. The five-planet model provides a significantly better fit to the observed RVs ($\Delta\chi_\mathrm{RV}^2\,{=}\,77$). The gray error bars in the bottom panel show the uncertainties after accounting for the ``jitter'' term described in Section~\ref{sec:five_pl_fits}.}
\label{fig:RV_fits}
\end{figure*}

\section{Analysis}
\label{sec:results}

\subsection{Four-planet TTV and RV fits}
\label{sec:stability}

Figure~\ref{fig:TTVs} shows the TTV data. Planets d and b did not exhibit large TTVs over the four-year observational baseline. Planet c showed TTVs with a readily detectable amplitude of about $30$\,minutes. Even without a detailed analysis, it seemed unlikely that planets d, b, or e could have induced such large TTVs, due to their relatively large separations from planet c and nonresonant orbital periods. Nonetheless, we began with a thorough investigation of models involving only the four known planets.

We started by placing upper limits on the planets' eccentricities by requiring the system to be long-term stable. Specifically, we carried out $N$-body simulations of Kepler-139 with properties drawn from the observed posteriors, using the \texttt{WHFast} integrator \citep{Wisdom&Holman1991, Rein&Tamayo2015} from the \texttt{REBOUND} package. We incrementally increased one planet's eccentricity, with the other planets placed on initially circular orbits, and monitored for Hill-sphere crossings during integrations spanning $10^9$ orbits of the innermost planet. By requiring that the system survives (i.e., all planets avoid Hill-sphere crossings) in ${>}\,50$\% of our simulations, we derived the following constraints: $e_d\,{\lesssim}\,0.35$, $e_b\,{\lesssim}\,0.30$, and $e_c\,{\lesssim}\,0.65$. In the TTV and RV fits described below, we placed uniform priors on the planets' eccentricities between zero and these upper limits.\footnote{To enforce a uniform eccentricity prior during MCMC sampling, we adopted the prior $p(h,\,k)\,{\propto}\,(h^2 + k^2)^{-1/2}$ on the eccentricity vector components $k$ and $h$.}

We fitted Kepler-139's TTVs and RVs using our \texttt{TTVFast}-based model and a Nelder-Mead optimizer. There were $5$ free parameters per planet and an RV offset parameter, for a total of $21$ free parameters. As a first guess, we used the RV-measured masses from \citet{Weiss2024} and set the eccentricities to zero. We determined the best four-planet model by minimizing the $\chi^2$ summed over the RVs and the transit times. The best-fit four-planet model we found has $\chi^2 = 1{,}057$, with $253$ degrees of freedom across the RV and TTV data sets. The TTVs predicted by the best-fit four-planet model are shown in the top row of Fig.~\ref{fig:TTVs}. As expected, this model captures some of the features in the noisy TTVs of planets d and b, but fails to describe planet c's larger TTVs.

The RV data and best-fit model are shown in Fig.~\ref{fig:RV_fits}. The four-planet model provides a moderate-quality fit to the observed RVs, although it leaves some unmodeled structure in the residuals, as shown in the middle panel. If the TTV data are disregarded, a better four-planet fit to the RVs can be obtained ($\Delta\chi_\mathrm{RV}^2\,{=}\,35$), explaining why \citet{Weiss2024} found the four-planet model to be satisfactory (see their Fig.~47).\footnote{\citet{Weiss2024} presented a periodogram of the RV residuals after subtracting the effects of the known planets. Although this periodogram has a peak at ${\sim}\,350$~days, it is not statistically significant, and there
are seven taller peaks.} However, the model obtained by fitting only the RVs also predicts ${\sim}\,60$-minute TTVs for planets d and b, which are confidently ruled out by the timing data (see Fig.~\ref{fig:TTVs}).

\subsection{Five-planet TTV and RV fits}
\label{sec:five_pl_fits}

The preceding results point toward the existence of a fifth planet in the Kepler-139 system. To investigate this possibility, we fitted models involving five planets: the three transiting planets (d, b, and c), the outer giant (e), and a new planet (f). Historically, it has proven challenging to determine the period and mass of a nontransiting planet using only TTVs \citep[see, e.g.,][]{Ballard2011, Boue2012, Nesvorny2014}. Fortunately, for Kepler-139, we have additional relevant information: (1) the TTVs of planets d and b are ${\lesssim}\,30$\,minutes, and (2) the RV data spanning a decade show significant deviations from the best-fit four planet model. Using all the available information, we were able to arrive at a well-constrained five-planet model, as described below.

The five-planet model had $26$ free parameters. To tackle the complex optimization problem, we carried out a ``parallel-tempered MCMC'' analysis, an extension of the MCMC method intended to improve performance when the posterior is multimodal. Parallel-tempered MCMC addresses this issue by simultaneously sampling from ``tempered'' versions of the posterior function, in which the contrast between good and bad models has been reduced, helping the sampler explore the full posterior. The tempered copies of the posteriors are sampled in parallel, with periodic swaps between the Markov chains. This way, the ``hot'' chains promise to find disparate peaks in the posterior function, and the ``cold'' chains seek to thoroughly explore each peak (see \citealt{Swendsen&Wang1986, Earl&Deem2005, Vousden2016} for more details). We used the \texttt{ptemcee} \citep{Vousden2016} extension of the \texttt{emcee} code with $20$ different energy levels and an adaptive temperature ladder (see \citealt{Blunt2020, Brandt2021, Canul2021} for other applications of \texttt{ptemcee}). Each tempered posterior was sampled using $250$ walkers for $250{,}000$ total steps (the first $50$\% of steps were discarded as burn-in). As usual, the coldest Markov chain was used for statistical inference, and we recorded its state every $25$ steps. The resulting posterior was well-converged; each parameter had an autocorrelation length of ${<}\,1{,}500$ steps. The walkers were initialized with broad initial conditions, including randomly sampled masses, eccentricities, and longitudes of pericenter. The orbital periods and initial mean longitudes of the transiting planets were selected based on their observed transit times, with some Gaussian noise added to ensure independence. The initial guess for planet f's period was drawn randomly from $18$ to $1180$~days, the full range of configurations that are Hill stable \citep{Gladman1993} with respect to planets b and e.

\begin{table}
\centering
\caption{Median values and $1$-$\sigma$ uncertainties from the five-planet fit to Kepler-139's TTVs and RVs. Orbital periods and mean longitudes are defined with respect to the reference epoch $T_\mathrm{epoch}\,{=}\,2{,}454{,}950$\,BJD.}
\begin{tabular}{cc}
 \hline
 Parameter & Value\\
 \hline
 $P_f$ [days] & $355^{+2}_{-2}$\\
 $m_f$ [$M_\oplus$] & $36^{+10}_{-10}$\\
 $e_f\cos(\varpi_f)$ & $-0.01^{+0.04}_{-0.05}$\\
 $e_f\sin(\varpi_f)$ & $0.08^{+0.04}_{-0.06}$\\
 $\lambda_f$ [deg] & $286^{+10}_{-11}$\\
 \hline
 $P_d$ [days] & $7.3053^{+0.0002}_{-0.0002}$\\
 $m_d$ [$M_\oplus$] & $2^{+2}_{-1}$\\
 $e_d\cos(\varpi_d)$ & $0.00^{+0.06}_{-0.06}$\\
 $e_d\sin(\varpi_d)$ & $-0.11^{+0.09}_{-0.07}$\\
 $\lambda_d$ [deg] & $82^{+7}_{-7}$\\
 \hline
 $P_b$ [days] & $15.7719^{+0.0007}_{-0.0005}$\\
 $m_b$ [$M_\oplus$] & $7^{+3}_{-3}$\\
 $e_b\cos(\varpi_b)$ & $0.14^{+0.03}_{-0.03}$\\
 $e_b\sin(\varpi_b)$ & $-0.19^{+0.06}_{-0.04}$\\
 $\lambda_b$ [deg] & $321^{+4}_{-4}$\\
 \hline
 $P_c$ [days] & $157.03^{+0.01}_{-0.01}$\\
 $m_c$ [$M_\oplus$] & $13^{+8}_{-7}$\\
 $e_c\cos(\varpi_c)$ & $-0.07^{+0.02}_{-0.02}$\\
 $e_c\sin(\varpi_c)$ & $-0.11^{+0.04}_{-0.04}$\\
 $\lambda_c$ [deg] & $343^{+3}_{-2}$\\
 \hline
 $P_e$ [days] & $1904^{+75}_{-72}$\\
 $m_e$ [$M_\oplus$] & $378^{+48}_{-39}$\\
 $e_e\cos(\varpi_e)$ & $-0.01^{+0.05}_{-0.07}$\\
 $e_e\sin(\varpi_e)$ & $0.04^{+0.07}_{-0.05}$\\
 $\lambda_e$ [deg] & $274^{+30}_{-32}$\\
 \hline
\end{tabular}
\label{table:bestfit_vals}
\end{table}

The parallel tempered MCMC analysis identified three distinct families of solutions that describe the TTVs and RVs moderately well, in which $P_f\,{\approx}\,354$\,days, $P_f\,{\approx}\,384$\,days, and $P_f\,{\approx}\,685$\,days. After Nelder-Mead optimization, the best-fit $\chi^2$ values for these models were $920$, $958$, and $974$, respectively. Thus, the $354$-day solution provides the best fit to the data, with a $\chi^2$ that is more than $35$ lower than the other two models. Most of the improvement comes from a better fit to the RV data. Approximating the posterior distribution as a multivariate Gaussian function, the Bayes factor is $\exp(\Delta \chi^2/2)\,{\gtrsim}\,10^{8}$, which we regard as decisive evidence in favor of the $354$-day period model. Thus, Kepler-139 appears to be a relatively rare case in which the period of a nontransiting planet is uniquely determined.

The best-fit five-planet model features a $45$-$M_\oplus$ planet on a $354$-day orbit. The period is $40$~days longer than the period corresponding to the 2:1 MMR with planet c. The five-planet model dramatically improves the fit to the observed TTVs relative to the four-planet model (see Fig.~\ref{fig:TTVs}; $\Delta\chi_\mathrm{TTV}^2\,{=}\,60$). The five-planet model also fits the RV data better, as illustrated in the bottom two panels of Fig.~\ref{fig:RV_fits} ($\Delta\chi_\mathrm{RV}^2\,{=}\,77$). In terms of the BIC, which takes into account the additional free parameters, the five-planet model is favored by $\Delta\mathrm{BIC}\,{=}\,109$, indicating an overwhelming preference. Altogether, the $\chi^2$ value of the best-fit model was $920$ with $248$ degrees of freedom. We attribute the discrepancy between the $\chi^2$ value and the number of degrees of freedom to underestimated uncertainties for some TTV and RV datapoints, a common occurrence (see, e.g., \citealt{Nesvorny2012, Nesvorny2014, Hadden&Lithwick2016}).

Hereafter, we refer to the new planet as Kepler-139f. To determine its parameters and their uncertainties as reliably as possible, we carried out another MCMC analysis in which we attempted to rectify the problem of underestimated observational uncertainties. We rejected 7 transit times that were ${>}\,4$-$\sigma$ outliers, which lowered the $\chi^2$ of the best-fit model to $610$. Then, we re-fitted the data with four ``jitter'' terms included in the likelihood function, three for the planets' transit times and one for the RVs. The maximum-likelihood solution has $\sigma_\mathrm{jit,\,TTV,\,d}\,{=}\,6.3$\,mins, $\sigma_\mathrm{jit,\,TTV,\,b}\,{=}\,3.4$\,mins, $\sigma_\mathrm{jit,\,TTV,\,c}\,{=}\,0.0$\,mins, and $\sigma_\mathrm{jit,\,RV}\,{=}\,2.9$\,m\,s$^{-1}$.\footnote{Inflating the uncertainties by the same amount during the model comparison step does not affect our conclusions. The best-fit model remains favored over the four-planet model by $\Delta \chi^2\,{=}\,78$ and over the other five-planet models by $\Delta \chi^2\,{>}\,15$.} With the inflated uncertainties, the best-fit model has a $\chi^2$ of $246$, in agreement with the $248$ degrees of freedom.

With the timing outliers removed, and the uncertainties inflated, we carried out another parallel-tempered MCMC analysis, initializing the walkers near the best-fit model parameters. We used $20$ temperatures and $250$ walkers, which were evolved for $100{,}000$ steps. We saved the state every $10$ steps and discarded the first $10$\% of the chain as burn-in. The resulting posterior was well-converged, with each parameter having an autocorrelation length much smaller than one $50$th the total number of steps. Median values and uncertainties are reported in Table~\ref{table:bestfit_vals}, and a corner plot that includes a subset of the parameters is included in Appendix~\ref{sec:MCMC_posterior}. Note that the marginalized posterior for the mass of planet f has a median ($36\,M_\oplus$) that is about 1-$\sigma$ smaller than the best-fit value reported above ($45\,M_\oplus$).

\begin{figure}
\centering
\includegraphics[width=0.475\textwidth]{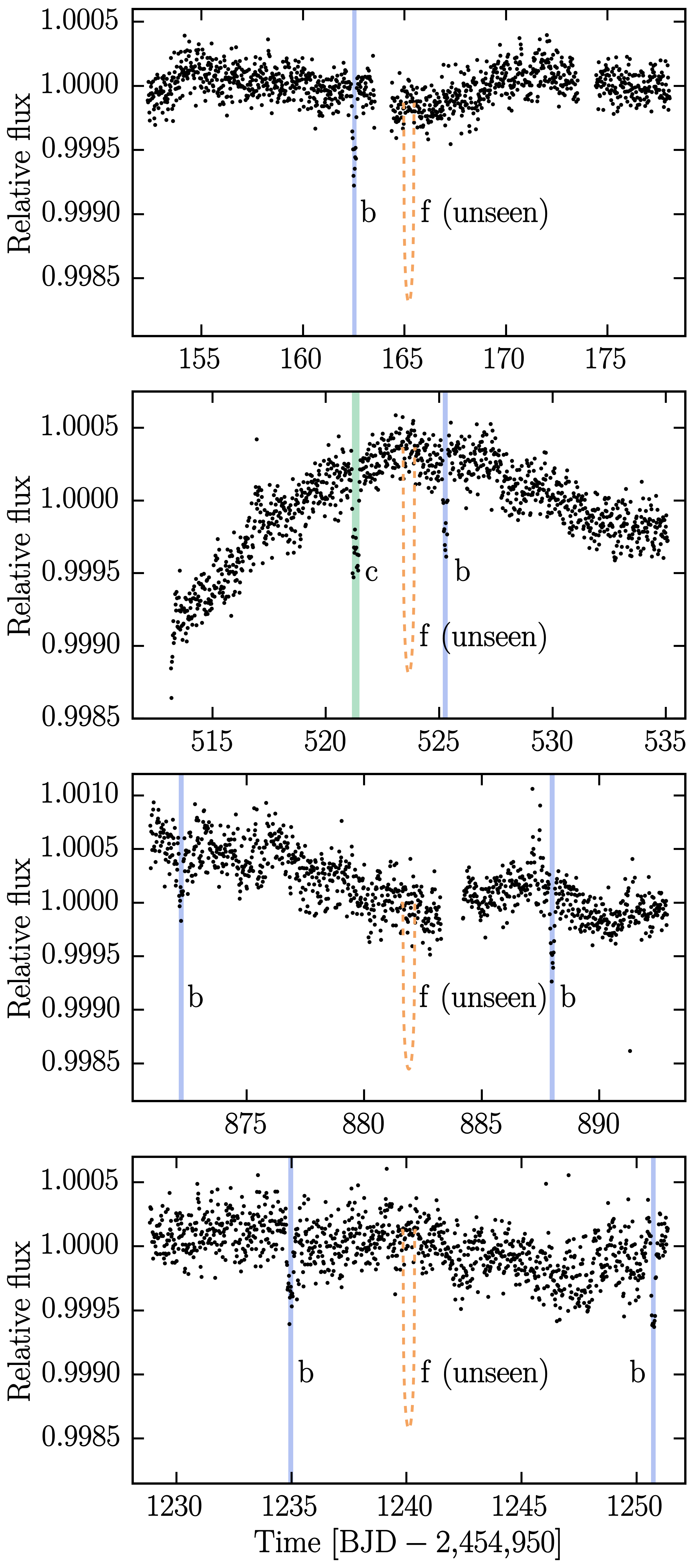}
\caption{{\it Kepler} observations of Kepler-139 over time ranges when transits of planet f would be expected. Each panel spans the $90$\%-confidence range of transit times predicted by the five-planet model (Table~\ref{table:bestfit_vals}). Transits of planets b and c are highlighted in blue and green. The expected transit signals for planet f, assuming $R_p\,{=}\,4\,R_\oplus$ and $b\,{=}\,0.5$, are shown in orange. Such signals would have been easily detectable but are not seen.}
\label{fig:raw_lightcurves}
\end{figure}

In addition to fitting the TTV and RV data well, the five-planet solution has other appealing qualities. Firstly, despite a broad prior on the planets' eccentricities, the model assigns small eccentricities (${\lesssim}\,0.1$) to the four inner planets, conforming to the typical pattern that has been observed in compact multiplanet systems \citep[e.g.,][]{VanEylen&Albrecht2015, Hadden&Lithwick2017}. Secondly, the model assigns comparable masses ($\sigma_m/\bar{m}\,{\approx}\,0.6$) to the three transiting planets, which is also a typical pattern \citep{Millholland2017} and consistent with the trends in the planets’ radii ($1.7\,R_\oplus$, $2.4\,R_\oplus$, and $2.5\,R_\oplus$ for planets d, b, and c, respectively). By contrast, in the four-planet model reported by \citet{Weiss2024} based on fitting RVs only, planet c had a large mass ($23\,{\pm}\,5\,M_\oplus$), corresponding to an unusually high mean density for a sub-Neptune-sized planet ($9\,{\pm}\,2$\,g\,cm$^{-3}$). With our updated mass measurements, planets d, b, and c have more typical densities of $2\,{\pm}\,2$\,g\,cm$^{-3}$, $3\,{\pm}\,1$\,g\,cm$^{-3}$, and $5\,{\pm}\,3$\,g\,cm$^{-3}$, respectively. The median densities increase with orbital period, although they are all consistent within their $1$-$\sigma$ uncertainties.

\begin{figure*}
\centering
\includegraphics[width=0.975\textwidth]{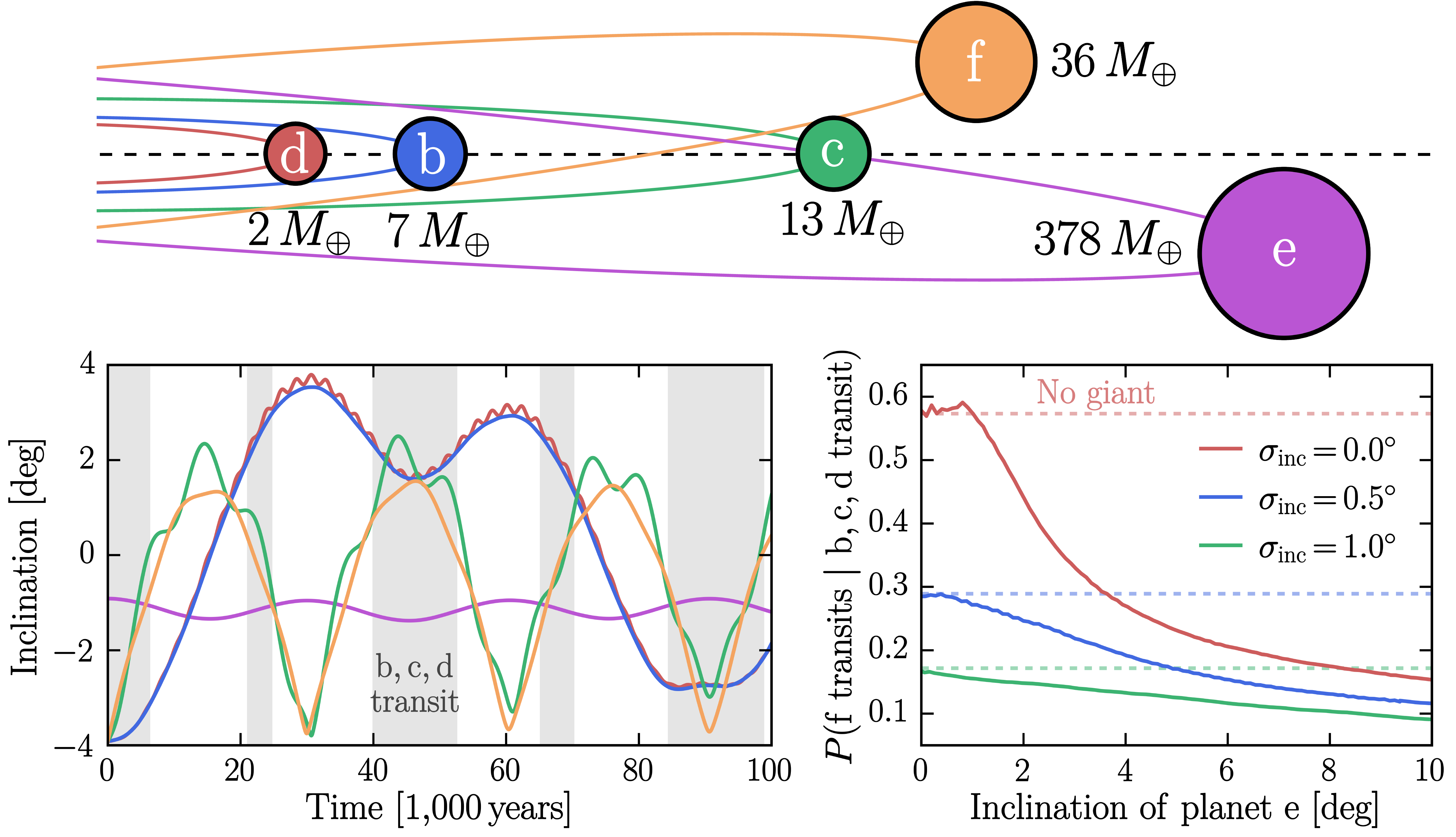}
\caption{Top: schematic diagram of the Kepler-139 system. The circles representing the planets are spaced logarithmically according to the planets' semi-major axes, with radii proportional to $\sqrt{R_p}$ (with $R_p$ estimated for planets f and e based on the mass-radius relation of \citealt{Chen2017}). Bottom left: secular (Laplace-Lagrange) evolution of the Kepler-139 planets' orbital inclinations, assuming the outer giant is inclined by $3^\circ$ with respect to the inner system. The gray time ranges are when planets d, b, and c have sufficiently low mutual inclinations to all be observable as transiting planets from a single line of sight (i.e., $P(\mathrm{b,c,d~transit})\,{>}\,0$). Bottom right: the time-averaged transit probability for planet f, given that planets d, b, and c transit, as a function of the outer giant's initial inclination. The dynamical influence of an inclined outer giant lowers the transit probability of planet f by a factor of a few. Dashed lines indicate the transit probability for planet f in the absence of an outer giant.}
\label{fig:LL_evol}
\end{figure*}

Any observed transits of Kepler-139f would have been detected easily by {\it Kepler}. With a mass of ${\sim}\,35\,M_\oplus$, we expect Kepler-139f to have a larger radius than planets d, b, and c. According to the probabilistic mass-radius relation of \citet{Chen2017}, Kepler-139f has a $90$\% chance of being larger than $4\,R_\oplus$ (${\sim}\,0.1$\% transit depth). Assuming an orbital period of $355$~days, planet f would have transited four times over the {\it Kepler} baseline, with a transit duration of ${\sim}\,12$~hours for an impact parameter of $0.5$. Figure~\ref{fig:raw_lightcurves} shows the portions of the {\it Kepler} light curve that span the $90$\%-confidence range of predicted transit times. A transit of a planet with radius $4\,R_\oplus$ or larger would have been readily detected but none are seen. We also searched other regions of the {\it Kepler} time series and did not find any transit-like events besides those of the known planets.

\section{Secular evolution and transit probabilities}
\label{sec:inc_sims}

We studied the dynamical influence of the outer giant planet on the inner planetary system by conducting dynamical simulations of the Kepler-139 system, adopting the parameters of the maximum-likelihood model (for which the predicted TTVs are RVs are shown in Fig.~\ref{fig:TTVs} and Fig.~\ref{fig:RV_fits}). The influence of a distant, massive perturber on smaller inner planets is often modeled with second-order secular (``Laplace-Lagrange'') theory \citep[e.g.,][]{Boue&Fabrycky2014, Lai&Pu2017, Becker&Adams2017}. The Laplace-Lagrange solution is derived by keeping the lowest-order secular terms in the planetary disturbing function, resulting in two independent sets of linear first-order differential equations that govern the planets' eccentricities and inclinations \citep{Murray&Dermott1999}. The accuracy of this approximation depends on the properties of the system, breaking down when planets are too closely spaced or are nearly resonant. We derived the Laplace-Lagrange equations for Kepler-139 with the help of the \texttt{LaplaceLagrangeSystem} class from the \texttt{celmech} package \citep{Hadden&Tamayo2022}. We tested the validity of the approximation by comparing some trial predictions of Laplace-Lagrange theory to those of $N$-body simulations performed with the \texttt{REBOUND} package. We found the approximation to hold well for Kepler-139, in that the frequencies and amplitudes of long-term inclination oscillations were accurate within ${\sim}\,10$\%.

Figure~\ref{fig:LL_evol} shows the Laplace-Lagrange evolution of the five Kepler-139 planets' inclinations over $100{,}000$\,years, assuming that the four inner planets are initially coplanar ($i\,{=}\,90^\circ$) and the outer giant is inclined by $3^\circ$ with respect to the inner system. The longitudes of the ascending node were drawn randomly from a uniform distribution between $0$ and $2\pi$. Because the innermost planets, d and b, are relatively closely spaced, they are tightly coupled and their inclinations remain the same to within a small fraction of a degree throughout the evolution (see the red and blue curves in Fig.~\ref{fig:LL_evol}). Likewise, the inclinations of planets c and f (the orange and green curves) track each other, although not as closely as those of planets d and b. Because of the relatively large gap between the orbits of planets b and c, the mutual inclination between the d/b pair and the c/f pair varies with an amplitude of nearly 8$^\circ$. Because the c/f pair is relatively close to the giant planet e, the inclination oscillations of the c/f pair and those of planet e are strongly correlated and 180$^\circ$ out of phase.

Could the inclination oscillations induced by the giant planet e be responsible for preventing planet f from transiting? To investigate this question, we used a Monte Carlo approach to calculate mutual transit probabilities under different assumptions about the initial orbital orientations (for alternative approaches, see \citealt{Ragozzine&Holman2010, Lissauer2011b, Brakensiek&Ragozzine2016}). We created many realizations of the Kepler-139 system, viewed them from random directions, and recorded which planets (if any) transit. For each realization, we chose the inclination $i_{\rm ref}$ of the system's reference plane with respect to the line of sight by drawing a value of $\cos i_{\rm ref}$ from a uniform distribution between $-1$ and $1$. Then, we drew initial orbital inclinations with respect to the reference plane from a Rayleigh distribution with a scale parameter $\sigma_\mathrm{inc}$. We conducted experiments with three different choices of $\sigma_\mathrm{inc}$:~$0.0^\circ,~0.5^\circ,~\mathrm{and}~1.0^\circ$. The longitudes of the ascending node, $\Omega_p$, were drawn randomly from a uniform distribution between $0$ and $2\pi$. A planet was considered to be transiting whenever $a_p \cos(i_{p,\mathrm{sky}})/R_\ast$ was between $-1$ and $1$, where
\begin{equation} 
\label{mutual_inc}
\cos i_{p,\mathrm{sky}}\,{=}\,\cos i_\mathrm{ref} \cos i_p \,{+}\,\sin i_\mathrm{ref} \sin i_p \cos \Omega_p.
\end{equation}
The gray regions in Fig.~\ref{fig:LL_evol} are the time ranges when there exist lines of sight from which d, b, and c are all transiting planets, as they are in reality. These gray regions occupy about $43$\% of the plotted timespan. Similarly, over about $42$\% of the timespan, there exists a line of sight in which planets d, b, and c transit but f does not, making this dynamical history plausible; of course, we may be observing Kepler-139 at a somewhat fortunate moment (see below).

The right panel of Fig.~\ref{fig:LL_evol} shows the probability that planet f transits, given that planets d, b, and c also transit, as a function of planet e's initial inclination. The plotted probability is an average over both viewing direction and time. For these calculations, we created $300{,}000$ realizations of the Kepler-139 system for each choice of planet e's inclination, and we recorded which planets transit over $10^6$\,years of Laplace-Lagrange evolution. When the outer giant is misaligned by a few degrees with respect to the inner system, secular evolution often prevents planet f from transiting even when planets d, b, and c are all transiting; specifically, as the outer giant's inclination is increased from $0^\circ$ to $10^\circ$, the conditional transit probability for planet f drops from $58$\% to $15$\% (assuming $\sigma_\mathrm{inc}\,{=}\,0.0^\circ$).

We also investigated a hypothetical system in which the outer giant planet does not exist at all. In that case, if the inner system were perfectly coplanar and planets d, b, and c transit, there is a $42$\% chance of observing the system from a line of sight in which planet f does not transit. For larger inclination dispersions, it becomes increasingly likely that planet f will avoid transiting, but also increasingly unlikely that planets d, b, and c will transit simultaneously. Quantitatively, increasing $\sigma_\mathrm{inc}$ from $0.0^\circ$ to $1.0^\circ$ lowers the probability that d, b, and c are all transiting planets from $0.83$\% to $0.57$\%.

Thus, we conclude that the existence of a slightly inclined outer giant planet reduces the transit probability for planet f by a factor of a few, relative to a hypothetical system in which the outer giant planet does not exist or is exactly aligned with the inner system. This makes it natural to suspect that the outer giant planet is the reason why planet f does not transit, although it is not possible to assign the blame with certainty.

A more subtle question is whether the observation that planets d, b, and c are all transiting can be used to place an upper limit on the mutual inclination of the giant planet. As the inclination of the outer giant is increased, secular evolution makes it less likely that planets d, b, and c will all be viewed as transiting planets. We found that the probability that d, b, and c all transit drops by a factor of ${\sim}\,3$ when the inclination of the outer giant is increased from $0^\circ$ to $5^\circ$. It is tempting to argue on this basis that the inclination of the outer giant is likely to be no more than a few degrees. Similar arguments have been presented for the multitransiting systems Kepler-129 \citep{Zhang2021} and HD~191939 \citep{Lubin2022}. However, such arguments are difficult to sustain for individual systems; Kepler-139 and the other systems were selected for study {\it because} they host multiple transiting planets, making it plausible that we are observing them at somewhat unusual moments in their dynamical evolution. One way to develop and test this argument is by considering population statistics and the observed transit multiplicity function, but this is beyond the scope of our study.

\section{Discussion and conclusion}
\label{sec:discussion}

Since the discovery that compact multiplanet systems are abundant, it has been appreciated that outer giant planets can significantly impact the formation and dynamics of these systems. In particular, outer giants are expected to perturb the orbits of their inner planets, reducing mutual transit probabilities. Consistent with this picture, we found evidence for a ${\sim}\,35$-$M_\oplus$ nontransiting planet, Kepler-139f, orbiting just outside of the previously known transiting inner planets. We showed that the system's outer giant induces inclination oscillations that make planet f less likely to transit simultaneously with the other planets, depending on the outer giant's inclination.

A related trend in the population of compact multiplanet systems was recently uncovered by \citet{Millholland2022}. They found statistical evidence that {\it Kepler} systems are ``truncated'' at periods of several hundred days, in the sense that the outermost known transiting planet tends to have a shorter period than one would expect based on selection effects alone. Specifically, \citet{Millholland2022} explored the detectability of a hypothetical outermost planet in many {\it Kepler} systems, whose properties were chosen based on the spacings and radii of the known planets, and reported that such a planet would have been detected in ${\sim}\,35$\% of systems. This suggests that either such outermost planets often do not form, or they differ meaningfully in their sizes, orbital separations, or mutual inclinations from their transiting neighbors.

Outer giant planets offer a promising possible explanation for this finding. First, we note that the occurrence rate of outer giants around inner systems is comparable to the fraction of systems that appear to be missing an outer planet \citep[e.g.,][]{Bryan2019, Rosenthal2022}. Second, the rarity of transiting outer giants suggests that their orbital planes are often at least slightly misaligned (by ${\gtrsim}\,3^\circ$) with respect to their inner systems \citep{Herman2019, Masuda2020}. Third, and most pertinent to this paper, it is easy for an outer giant planet to lower the transit probability of a planet on the outer edge of an inner system. This is because (1) the inner planets that are closest to the outer giant are dynamically perturbed most strongly by the giant, and (2) at wider orbital separations, even a small mutual inclination is enough to deny a planet the possibility of transiting along with its inner companions. Kepler-139 provides an illustrative example of these effects: an inclined outer giant makes Kepler-139f unlikely to transit even if the inner system was initially exactly coplanar.

\citet{Millholland2022} and \citet{Sobski&Millholland2023} proposed, and ultimately argued against, a more dramatic version of this hypothesis, in which outer giants truncate their inner systems by scattering or ejecting the putative outermost planets. Destabilizing the inner systems requires close-in giants, which they argued would have been detected in many systems. By contrast, even a distant outer giant can easily excite the necessary mutual inclinations of planets on the outer edges of inner systems.

This hypothesis can be tested by searching for more examples of nontransiting planets in systems known to harbor outer giant planets. For Kepler-139, the TTV information was crucial to our discovery. The RV data alone were not conclusive, but were crucial in narrowing down the possibilities for the period and mass of the perturbing planet. Thus, we encourage further RV monitoring and transit timing of compact multiplanet systems as a means of bringing the connection between inner and outer planetary systems into sharper focus.

\section{Acknowledgments} 
\label{sec:acknowledgments}

We are grateful to the KGPS team for making their rich RV dataset publicly available. We thank the anonymous referee for a timely and thoughtful report, as well as Jeremy Goodman for useful discussions. We are pleased to acknowledge that the work reported in this paper was substantially performed using the Princeton Research Computing resources at Princeton University, which is a consortium of groups led by the Princeton Institute for Computational Science and Engineering (PICSciE) and the Office of Information Technology’s Research Computing.

\appendix

\section{MCMC posterior for five-planet model}
\label{sec:MCMC_posterior}

\begin{figure*}
\centering
\includegraphics[width=0.95\textwidth]{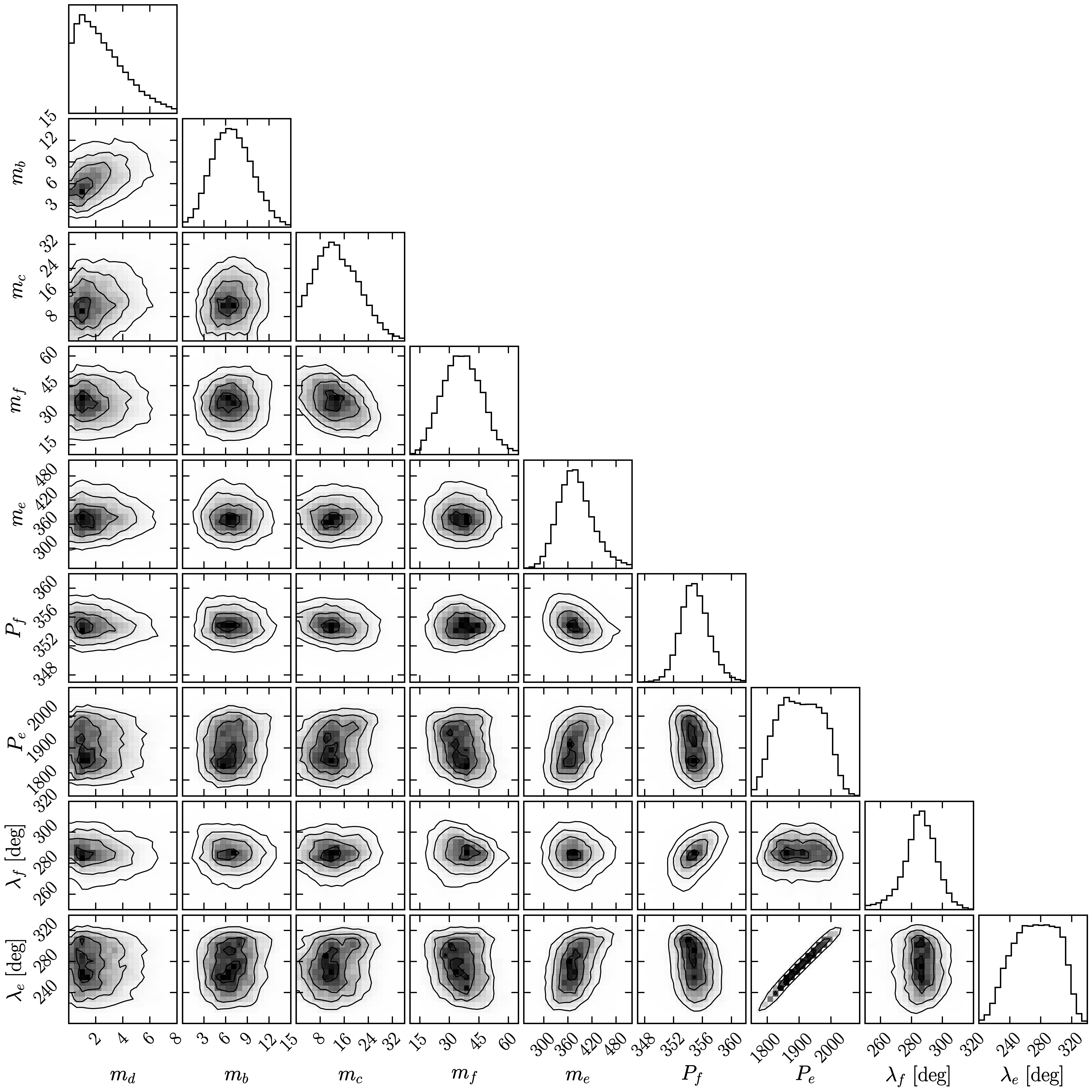}
\caption{Posterior probability densities for the parameters of the five-planet model for Kepler-139, fitted to the TTVs and RVs. Black contours denote the $0.5\sigma$, $1.0\sigma$, $1.5\sigma$, and $2.0\sigma$ joint confidence levels, and $1$D histograms show the marginalized distributions. To make this figure of manageable size, we omitted the eccentricity parameters and the periods and mean longitudes of the three transiting planets; the median values and uncertainties for those parameters are reported in Table~\ref{table:bestfit_vals}.}
\label{fig:corner}
\end{figure*}

Figure~\ref{fig:corner} is a ``corner plot'' that displays the joint posterior probability densities based on fitting the five-planet model to the available TTVs and RVs of Kepler-139. The posterior is complex, but nonetheless, most parameters are well constrained, with approximately Gaussian distributions. There are some degeneracies, primarily between the planet's orbital periods and their eccentricities or initial mean longitudes.

\bibliography{refs}{}
\bibliographystyle{aasjournal}

\end{document}